# Understanding Groups' Properties as a Means of Improving Collaborative Search Systems

Meredith Ringel Morris, Jaime Teevan
Microsoft Research
Redmond, WA, USA
{merrie, teevan}@microsoft.com

## ABSTRACT
Understanding the similar properties of people involved in group search sessions has the potential to significantly improve collaborative search systems; such systems could be enhanced by information retrieval algorithms and user interface modifications that take advantage of important properties, for example by re-ordering search results using information from group members' combined user profiles. Understanding what makes group members similar can also assist with the identification of groups, which can be valuable for connecting users with others with whom they might undertake a collaborative search. In this workshop paper, we describe our current research efforts towards studying the properties of a variety of group types. We discuss properties of groups that may be relevant to designers of collaborative search systems, and propose ways in which understanding such properties could influence the design of interfaces and algorithms for collaborative Web search.

## Keywords
Collaborative search, collaborative information retrieval, Web search, personalization, groupization.

## 1. INTRODUCTION
Recently, researchers have begun to introduce systems that facilitate collaboration amongst groups of users on Web search tasks (*e.g.*, [1], [4], [6]). Our research aims to understand the properties associated with groups of people with common search goals in order to facilitate the design of algorithms, interactions, and interfaces that can leverage those properties to enhance collaborative Web search. Our research is informed by work on personalization [8], which has used properties of an individual (*e.g.*, a user's past browsing history, past query history, or desktop content) to enhance that individual's Web search experience; we aim to provide analogous support for collaborative Web search by identifying relevant properties of groups of users.

## 2. GROUP TYPES
We are studying groups along two axes as part of our research effort. The first axis relates to the longevity of group membership: groups can either be short term (*task-based*) or long term (*trait-based*). The second axis relates to how group membership is determined: it can be determined either by information provided by group members (*explicit*) or inferred from member activity (*implicit*). Although this workshop is primarily concerned with issues relating to explicit, task-based groups, we have included some discussion of the other types of groups in this position paper, as lessons learned from explicit, task-based groups may benefit them, or vice-versa.

### 2.1 Group Longevity
Short-term groups are comprised of people with a shared goal. Group members are working together to accomplish this shared task; hence, we refer to these groups as *task-based* groups. Common tasks that may motivate groups to collaborate on Web search include travel planning, shopping, work- or school-related projects and reports, social planning, or medical searches [5].

Long-term groups are comprised of users who are related through shared traits or long-term interests. We refer to these groups as *trait-based* groups. Group members may not be consciously collaborating on the same task, but may be highly likely to repeat or augment tasks already accomplished by other group members, have interests in the same queries and results as other group members, and/or possess information relevant to another group member's task. We are studying a variety of trait-based groups built from shared interests, occupations, geography, and/or demographics.

Interest groups are comprised of users with shared interest in a particular topic. We have been using e-mail distribution lists as one means of studying interest groups (*i.e.*, users who have subscribed to email discussion groups on topics such as photography, pets, or vegetarianism). Another approach we have used to group people by interest is to compare the similarity of the content on an individual's computer to that on others' computers.

Occupational groups are comprised of users with related jobs. We have explored two main classes of occupational groups. *Job-role* groups consist of people with similar job titles who may work on different products or even in different companies (*i.e.*, a group of software engineers or a group of pediatricians). *Job-team* groups consist of people who work on the same product or for the same company; such groups may consist of people with heterogeneous job roles (*i.e.*, the team of people that works on Microsoft Word).

Geographic groups are comprised of users who live or work in a particular region. This relationship can be hierarchical (*i.e.*, a group could be based on city, county, state, country, etc.). Demographic groups are based on users who share characteristics such as gender or age.

### 2.2 Group Identification
The other axis we have explored is how groups are identified. One way group membership can be determined is by information provided directly from the members. We consider these groups *explicit* groups. For example, an explicit task-based group is one where group members are overtly collaborating on a specific task. Group membership can also be inferred. We call these groups *implicit* groups. An implicit task-based group may be formed from people who appear, based on their actions, to be conducting the same task.

We have explored several methods of group identification with explicit data, including explicit task-based collaboration, mailing list membership, gender, age, geographic location, job-role, and job-team. We have also explored several methods for implicit group identification, including grouping users with similar desktop indices, grouping users who issue similar search queries, and grouping users with similarities in relevance judgments.

The ability to identify groups "in the wild" is important to the success of groupization techniques. However, we believe that this is an attainable goal; here, we propose techniques for identifying our target group types.

Explicit, task-based groups can be identified through their use of a collaborative search tool, such as SearchTogether, a freely available Internet Explorer plug-in for collaborative searching [http://research.microsoft.com/searchtogether/].

Explicit, trait-based information can be gathered from profiles that some (though not all) users fill out in order to register with and access custom features of many popular search engines. E-mail and instant-messaging contacts could also be used to construct group membership information. Additionally, collaborative search tools for the enterprise, for use on corporate intranets, would likely have access to employee directories with a variety of demographic information including job titles and hierarchies.

Implicit groups can be identified via the use of server-side metrics that many search companies typically gather (in some cases only for users who have opted in to special services such as search toolbars, query histories, or personalization). Geographic data can be gleaned from IP addresses [3], and it may be possible to infer gender from query history [2]. Use of special-topic websites or portals may indicate interest-based groups [7]. One can imagine that collaborative search tools might evolve as part of social networking sites (*e.g.*, Facebook, MySpace) in which users' profiles and network structures could provide rich information relevant to several trait-based grouping categories. Relevance-judgment similarity may be able to be approximated using click-through data, or using data from social bookmarking tools such as del.icio.us [http://del.icio.us/].

## 3. GATHERING GROUP DATA

We have conducted two experiments to gather data that can be used to study what makes group members similar and inform enhancements for group Web search. In this section, we briefly describe the methodologies used and types of data gathered.

Both studies involved gathering implicit and explicit properties about each participant that could be used to determine group membership, as well as gathering each participant's explicit relevance judgments for a number of queries. This was done using software developed by our research group. Users were asked to select queries from a list of queries. Selecting a query displayed a list of pre-cached search results, ordered randomly so that participants' judgments were not biased by rank. All results were displayed in standard title + snippet + URL format; clicking on a result opened the target page in a new browser window. Next to each result were radio buttons that allowed the participant to mark whether they deemed that result to be highly relevant, relevant, or not relevant to the current query. Our software also reported scores for how well each result fared according to various personalization metrics [8], such as whether the target URL has been previously visited or bookmarked by the participant, and how frequently the terms in the result appeared in the participant's desktop index.

### 3.1 Study 1: Trait-Based Data

For the first study, 110 participants, all Microsoft employees, were recruited via advertisement to some of the company's interest-based e-mail lists (pet ownership, photography enthusiasts, and vegetarians) and advertisement to members of specific teams within the company (two product groups and one research group). Participants provided additional trait-based group membership information, such as their gender, age, residential ZIP code, and job title.

Participants completed the study from their own computers. They were instructed to choose any three of six pre-defined socially-themed queries and any three of six pre-defined work-themed queries. Participants provided relevance judgments for the top 40 search results for each of their six chosen queries, using the study software described above.

### 3.2 Study 2: Task-Based Data

Thirty participants, all Microsoft employees, volunteered for the second study in task-based groups of three people each (for a total of ten groups). Group members all knew each other prior to the study, and had work-related tasks that resulted in a shared information need that they wished to address through Web search. Each group provided a brief description of its shared task, and each group member individually provided six queries that they might use to pursue the group's shared task. Participants also provided demographic information, such as age and gender, and descriptions of their expertise relevant to the group's task.

Participants then used our relevance-judgment software to provide judgments for 20 search results for each of 15 queries: 6 queries relevant to the group's shared task (using two of the queries suggested by each group member) and 9 "common" queries that all 30 study participants evaluated. These common queries were gathered by having a set of 3 judges rate which query from each group's task-specific set was most understandable to someone external to the group; this information was used to select one query from each of the 10 groups' sets that was evaluated by all of the other groups in the study.

## 4. ANALYZING GROUP DATA

There are two main classes of analyses that we are interested in carrying out on our data sets. First, we are interested in identifying which types of groups are likely to benefit from collaborative search enhancements: task- or trait-based, explicit or implicit. Are such techniques more effective for groups working on particular kinds of tasks or defined by particular types of traits?

Understanding the properties that are necessary to gather to identify useful groups, and whether they can be gathered implicitly or explicitly, will suggest where the opportunities for developing algorithms and interfaces to improve the group search experience lie. We are also interested in analyzing our data to evaluate the potential utility of various techniques to enhance collaborative Web search.

### 4.1 Properties of Groups

We are in the process of exploring whether groups determined by the different properties described in Section 2 are comprised of members that are similar to each other along several dimensions, including the queries the group members generated or selected, the relevance judgments they provided for search results, and the user profile information we collected.

Query choice is one property our data enables us to explore from several perspectives. For example, for Study 1 we can explore whether participants' choice of which socially-themed and/or work-themed queries to evaluate predicts group membership. For Study 2, we can examine the degree of overlap in group members' query suggestions.

We are also interested in whether similarities in relevance judgment patterns can be used to predict group membership, and how judgment similarities may differ based on factors such as query category (*i.e.*, social vs. work queries in Study 1) or query origin (*i.e.* on-task vs. common queries in Study 2).

The variation (or lack thereof) amongst user profiles for group members is another area of interest; user profiles can be represented as a vector of term frequencies from a user's desktop index. Are the profiles of group members more similar to each other than those of non-group members? Are there interesting variations amongst group members' profiles that could be utilized by collaborative search systems?

## 4.2 Enhancements for Groups

Based on preliminary analyses of our studies' data, we have identified several possible algorithmic and user interface enhancements for collaborative search systems. Preliminary evaluations of the utility of these concepts are possible from our existing data sets; should they prove promising, evaluation in the context of a prototype collaborative search system would be a valuable next-step.

One potential enhancement to collaborative search systems is the extension of personalization algorithms to take advantage of data from all group members; we refer to this process as *groupization*. For example, personalization scores from all group members could be combined arithmetically to re-rank results lists in a manner which reflects the group's shared interests and expertise. A "groupized" list might be appropriate for display in a collaborative search tool. Groupization may also be useful even in single user scenarios, where data from a user's past collaborators, trait-based group members, or implicit task-based group members could be used to perform groupization, which may be beneficial since many personalization algorithms work better when there is increased data available [8].

Knowledge of a group's properties could be used to improve division of labor amongst collaborators. For example, SearchTogether [4] provides a "split search" feature, in which one group member types a query and the results are divided round-robin style amongst the collaborators. By comparing the relative similarity of each result to each group member's desktop index, one could potentially optimize a collaborative search tool so that it delivers to each group member the subset of results most relevant to that individual's unique expertise.

Group members' query terms could also be used to provide a better collaborative search experience. For instance, results that contain terms not just from the current query, but from other group members' recent, related queries, could be ranked more highly. In addition to re-ranking results based on other group members' on-task queries, a user's query could be modified (perhaps invisibly) by a collaborative search system, such as by appending other group members' previously issued queries using operators such as "prefer:". The user interface for displaying search results could also call out results that seem relevant to multiple group members' queries, perhaps through variations on traditional hit-highlighting techniques.

Although current collaborative search prototypes, such as [1], [4], or [6], are designed for use by task-based groups of users, it may be the case that some users could benefit from collaborative search (*i.e.,* for increased coverage of relevant results, or simply for the social pleasure of collaboration), but are not members of a task-based group. The ability to identify potential task-based groups, perhaps based on membership in trait-based groups or based on properties such as query similarity, could be used to serendipitously match people with potential search partners with whom they could enter into explicit collaborations.

## 5. CONCLUSION

The design of collaborative search systems can benefit from reflecting on single-user techniques, such as personalization, and considering how they might be applied to groups. Similarly, single-user search tools might benefit from knowledge derived from collaborative systems; for example, collaborative systems may incent users to provide explicit relevance judgments on content (for use by their collaborators) which could then be utilized by search engines to revise result rankings that are displayed to users in traditional, single-user search scenarios.

We are interested in studying the properties of a variety of groups, spanning the spectrum from implicit to explicit and from trait-based to task-based, and considering how the lessons learned could be used to facilitate a fruitful collaborative search experience.

## 6. ACKNOWLEDGMENTS

We gratefully acknowledge Steve Bush for his work developing and deploying software for gathering data related to these efforts. We also acknowledge Susan Dumais for many insightful discussions related to these issues.